\documentclass[amsmath,amssymb,prl,twocolumn]{revtex4}

\usepackage{graphicx}

\begin{document}

\title{Phonons and $d$-wave pairing in the two-dimensional Hubbard model}
\author{Carsten Honerkamp$^{1,2}$, Henry C. Fu$^3$, and Dung-Hai Lee$^{3,4,5}$}
\affiliation{$^1$Theoretical Physics, Universit\"at W\"urzburg, D-97074 W\"urzburg, Germany \\
$^2$ Max-Planck-Institute for Solid State Research, D-70569 Stuttgart, Germany \\
$^3$ Department of Physics, University of California at Berkeley, CA-94720
Berkeley, USA \\
$^4$ Center for Advanced Study, Tsinghua University, Beijing 100084, China \\
$^5$ Material Science Division, Lawrence Berkeley National
Laboratory,Berkeley, CA 94720, USA. 
}

\date{May 5, 2006}

\begin{abstract}
We analyze the influence of phonons on the $d_{x^2-y^2}$-pairing instability in the
  Hubbard model on the two-dimensional square lattice at weak to
moderate interaction $U$, using a functional renormalization group scheme with
frequency-dependent interaction vertices. 
As measured by the pairing scale, the $B_{1g}$ buckling mode enhances the pairing, while other phonon modes decrease the pairing. 
When various phonon modes are included together, the net effect on the scale is small. 
However, in situations
where $d$-wave superconductivity and other tendencies, e.g. antiferromagnetism, are closely competing, the combined effect of different phonons may be able to tip the balance towards pairing.

\end{abstract}


\maketitle
The two-dimensional Hubbard model is one of the most-studied models in context with high-temperature superconductivity in the layered cuprates. 
An important question is whether electronic interactions of the Hubbard model alone can provide a sufficient pairing strength explaining the high critical temperatures and large energy gaps observed experimentally. Furthermore, in particular for large values of the Hubbard interactions, there is a strong competition between various ordered states such that one may want to have an argument why superconductivity prevails in a large parameter region for most of the materials. 
Although the lattice degrees of freedom were thought to be irrelevant for the high-$T_c$ problem for a long time\cite{anderson}, and phononic signatures in the electronic properties are still debated intensively\cite{kink}, an additional phononic contribution to the pairing seems to be a natural way to enhance the superconducting pairing against other competing electronic correlations. 

A theoretical analysis of this question in the Hubbard model at large values of the onsite interaction is difficult. The impact of phonons on the pairing interaction has been addressed in various ways with partially contradicting results\cite{huang,zeyher,grilli,entel}. Here we propose to analyze the situation at weak to moderate values of the coupling constant. This will allow us to obtain some qualitative insights. The method we use is the functional renormalization group (fRG). Previously, this method has been used to classify the leading instabilities of the weakly coupled Hubbard model without phonons\cite{zanchi,halboth,hsfr,tsai}. There, for band fillings when the Fermi surface (FS) is not nested, a $d_{x^2-y^2}$-wave superconducting instability is obtained in a large parameter window. The main driving force for these superconducting tendencies are antiferromagnetic (AF) spin fluctuations. Here we add phonon-mediated interactions to the bare Hamiltonian. We analyze the changes in the critical energy scale for the Cooper instability and in the competition with other states.

The fRG scheme we use is an approximation to an exact flow equation for the one-particle irreducible vertex functions of a many-fermion system\cite{fRG}. The quadratic part of the fermionic action is supplemented with a cutoff function which restricts the functional integral over the fermions to the modes with dispersion  $|\xi (\vec{k})|> \Lambda$. For the 2D square lattice we use a $t$-$t'$-parameterization, $\xi (\vec{k} )= -2t (\cos k_x +\cos k_y)- 4t' \cos k_x \cos k_y -\mu$ with nearest and next-nearest neighbor hoppings $t$ and $t'$ and chemical potential $\mu$.
The fRG flow is generated by lowering the RG scale $\Lambda $ from an initial value $\Lambda_0 \sim$ bandwidth. Thereby momentum shells with energy distance $\Lambda$ to the FS are integrated out. In the approximation we use, the change of the interaction vertex is given by one-loop particle-hole (including vertex corrections and screening) and particle-particle pairs where one intermediate particle is at the RG scale $\Lambda$ while the second one has $|\xi (\vec{k})|\ge  \Lambda$. Higher loop contributions are generated by the integration of the RG flow. The flow of the self energy will be analyzed in a later publication (see also \cite{katanin}), and as in previous studies its feedback on the flow of the interaction vertex is neglected. \\
The interaction vertex $V_\Lambda (k_1,k_2,k_3)$ depends on the generalized wavevectors of two incoming particles ($k_1$ and $k_2$) and one outgoing ($k_3$) particle with wavevector, Matsubara frequencies and spin projection $k_i=(\vec{k}_i, \omega_i,s_i)$. The dependence of this function on three wave-vectors is discretized in the so-called $N$-patch scheme, introduced in this context by Zanchi and Schulz\cite{zanchi}. This scheme takes advantage of the tested fact that for standard Fermi liquid instabilities the leading flow is rendered correctly by projecting the wavevectors $\vec{k}_1,\vec{k}_2$ and $\vec{k}_3$ on the FS and keeping the variation of $V_\Lambda(\vec{k}^F_1,\vec{k}^F_2,\vec{k}^F_3, \dots)$ when the $\vec{k}^F_i$ are varied around the FS. 
 Hence, one calculates $V_\Lambda(\vec{k}_1,\vec{k}_2,\vec{k}_3, \dots)$  for $\vec{k}_1,\vec{k}_2,\vec{k}_3$ on the FS and treats it as piecewise constant when $\vec{k}_1,\vec{k}_2$, and $\vec{k}_3$ move within elongated patches stretching from the origin of the Brillouin zone (BZ) to the $(\pm \pi,\pm \pi)$-points.
In order to treat retarded interactions we have to go beyond the previous works which neglected the frequency dependence\cite{tam}. We divide the Matsubara frequency axis into $M$ sections. The aim is to approximate the decay of a phonon propagator above a characteristic frequency $\omega_0$. Below we show results for 32 BZ patches and $M=10$. The minimal frequency spacing is $\omega_0 = 0.2t$. The frequencies for which the vertices are computed range between $\pm 5 \omega_0 $. 
The frequencies of the dispersion-less phonons considered here are taken to be less than $\omega_0$. We have checked that other reasonable choices do not change our qualitative findings. 

The RG flow is started at an initial scale $\Lambda_0$ with initial interaction $V_\Lambda (\vec{k}_1,\vec{k}_2,\vec{k}_3, \omega_1, \omega_2, \omega_3)$.
What is typically encountered at low $T$ is a {\em flow to strong coupling}, where for a certain flow parameter $\Lambda_c$ one or several components of $V_\Lambda (\vec{k}_1,\vec{k}_2,\vec{k}_3,\omega_1, \omega_2, \omega_3)$ become large. At this point the approximations break down, and the flow has to be stopped. Physical information about the low-energy state is obtained by analyzing which coupling functions and susceptibilities grow most strongly. 
For standard Cooper instabilities, the critical scale $\Lambda_c$ at $T=0$ is proportional to the critical temperature $T_c$. 

For pure Hubbard interactions, the initial vertex  at scale $\Lambda_0$ is $V_{\Lambda_0}^U(k_1,k_2,k_3) = U$. For phonon-mediated interactions, we add a retarded part, leading to 
\begin{eqnarray}
V_{\Lambda_0} (k_1,k_2,k_3) &=&  U -  \sum_{i} \frac{g_i(\vec{k}_1,\vec{k}_3)  g_i (\vec{k}_2,\vec{k}_4 ) 
\omega_{i,0}}{(\omega_1- \omega_3)^2 +\omega_{0,i}^{2}} \, . \label{phomedint}
\end{eqnarray}
Here,  $g_i(\vec{k}_1,\vec{k}_3)$ is the electron-phonon interaction for an (Einstein) phonon mode $i$ with frequency $\omega_0^i$ scattering an electron from $k_1$ to $k_3$. Obviously, if the product of the $g$'s in the numerator does not produce sign changes, the main effect of the phonon part is to reduce the effective onsite interaction $U$. The coupling strength of the mode can be measured by the FS average $\lambda_i = 2 \sum_{\vec{k},\vec{k}'} \delta (\xi_{\vec{k}}) \delta (\xi_{\vec{k}'}) \,  |g_i(\vec{k}, \vec{k}')|^2 / [ \omega_{0,i} \mathrm{Vol} \sum_{\vec{k}} \delta (\xi_{\vec{k}})] $.
Motivated by current issues in the high-$T_c$ problem, we analyze various different phonon modes, idealized as dispersionless. 
First we consider a Holstein phonon with a $\vec{k}$-independent coupling $g_{\mathrm{Holstein}}(\vec{k},\vec{k}')=g$. As a next step we analyze a coupling which only depends on the transferred wave-vector $\vec{q}=\vec{k}-\vec{k}'$, 
\begin{equation} |g_{\mathrm{Buck}} (\vec{q}) |^2 = g^2_{\mathrm{Buck}} \left( \cos^2 \frac{q_x}{2} + \cos^2 \frac{q_y}{2} \right)\, . \label{buck} \end{equation}  
A coupling of this type was used by Bulut and Scalapino\cite{bulut} in their analysis of the out-of-plane motion of the planar oxygens (buckling mode) in the language of a one-band Hubbard model. They and various other authors\cite{dahm,nazarenko,jepsen} pointed out that the suppression of this coupling for large momentum transfers $q\sim (\pi,\pi)$ leads to an attractive $d$-wave Cooper pairing potential. The same mode was also considered using multiband models involving the planar or full copper-oxide structure\cite{devereaux,jepsen}. Then the out-of-phase $c$-axis motion of the planar oxygens on $x$- and $y$-bonds, known as $B_{1g}$-mode, gives rise to a coupling with a sign change under 90 degree rotations, 
\begin{eqnarray} g_{\mathrm{B1g}} (\vec{k},\vec{k}')&=& g_{\mathrm{B1g}} \, \left[ \sqrt{ (1+\cos k_x) (1+ \cos k_x')} \right. \quad \nonumber \\ && \quad \left. - \sqrt{ (1+\cos k_y) (1+ \cos k_y')} \right]  \, .\label{b1g} \end{eqnarray} 
In optimally doped cuprates this mode shows up at $\sim 36$meV and its coupling strength is taken as $\lambda_{\mathrm{B1g}} = 0.23$\cite{devereaux}. 
In-plane breathing modes have also been observed and discussed in the cuprates. Here the planar oxygen atoms next to a copper site move towards or away from the copper atom. One obtains a coupling (again with $\vec{q}=\vec{k}-\vec{k}'$) $
|g_{\mathrm{breathe}} (\vec{q})|^2 =  g^2_{\mathrm{breathe}} \, \left( \sin^2 \frac{q_x}{2} + \sin^2 \frac{q_y}{2} \right) \, $\cite{bulut}.
The frequency for this mode in optimally doped samples is taken as 70meV, and the FS average of the coupling strength is quoted as $\lambda_{\mathrm{breathe}}=0.02$\cite{devereaux}. Finally there is a $c$-axis vibration of the apex oxygen above or below the copper with a coupling 
$ g_{\mathrm{apex}} (\vec{k},\vec{k}')= g_{\mathrm{apex}} \, \left( \cos k_x  - \cos k_y \right)\cdot \left( \cos k_x' -\cos k_y' \right) \, . $
The bare $ g_{\mathrm{apex}} (\vec{k},\vec{k}')$  does not support an attractive $d$-wave pairing component, but it has been argued\cite{devereaux_new} that screening can generate a $d$-wave attraction. The frequency of this mode in the cuprates is roughly 60meV, and the coupling strength is discussed to be as high as $\lambda_{\mathrm{apex}} = 0.5$ and higher\cite{devereaux_new}. Below we use the conversion 50meV$=0.1t$.

\begin{figure}

\includegraphics[width=.48\textwidth]{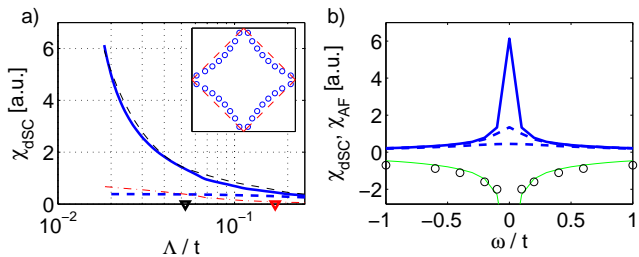}

\caption{(color online). a) RG flow of the susceptibilities $\chi_{\mathrm{dw}} (\omega)$ (thick (dashed) line for $\omega=0$ ($\omega=0.5t$)) and $\chi_{\mathrm{AF}}(\vec{q},\omega=0)$  at wavevector $\vec{q}=(\pi,\pi)$ (thin solid line) for $t'=-0.25t$, $U=2.5t$ and $\mu=-0.94t$ with $N=32$ FS points, $T=0.01t$. The marks on the $\Lambda$-axis denote where the largest coupling reaches $5t$ and $10t$. 
b) Frequency dependence of $\chi_{\mathrm{dSC}} (\omega)$ at the scale where $V_{\mathrm{max}}$ reaches $32t$ (thick solid line), $10t$ and $5t$ (dashed lines). The thin solid line shows the rescaled $\chi_{\mathrm{AF}} (\omega)$ at $V_{\mathrm{max}}=32t$. The open  circles show he FS-averaged $d$-wave pair scattering $V^{dw}_\Lambda (\omega,- \omega,\omega_m)$ versus transfered frequency $\omega$ (with outgoing frequency $\omega_m =\pm 0.1t$).
 } 
\label{susplot}
\end{figure}

First we describe the flow without phonons. We choose a curved FS near
the van Hove points with $n=0.83$ particles per site. Note that the
doping dependence at weak coupling is quite different from the
behavior at large $U$, so no strong conclusions about the doping
dependence in the cuprates can be drawn. There are two main 
effects of the interaction: tendencies towards AF spin-density wave (SDW) and
  towards $d$-wave pairing. The FS with the 32 discretization points, and the flows of $d$-wave pairing and AF-SDW susceptibility $\chi_{dw}$ and $\chi_{AF}$ are shown in Fig. \ref{susplot}. 
As in previous studies without frequency dependence, for these parameters $\chi_{dw}$ grows most strongly toward low scales, but one clearly observes tight competition with antiferromagnetism. Therefore, for these situation with the FS near the saddle points, alternative interpretations of this multi-channel instability have been considered\cite{hsfr}. Reducing $U$ or the nesting makes the $d$-wave pairing more dominant.  
In the right panel of Fig. \ref{susplot} we display the frequency dependence of $\chi_{dw}$ which shows the build-up of a zero-frequency peak when the instability is approached at low scales. We also plot the FS-averaged $d$-wave pairing interaction with zero total incoming frequency versus the frequency transfer. The peak is quite broad, signaling a large pairing `Debye'  energy scale $\omega_D \sim t$.  It roughly tracks the frequency dependence of $\chi_{AF}$ at wavevector $(\pi,\pi)$. A similar behavior has been found in dynamical cluster approximation (DCA) calculations\cite{maierfreq}. 

\begin{figure}

\includegraphics[width=.48\textwidth]{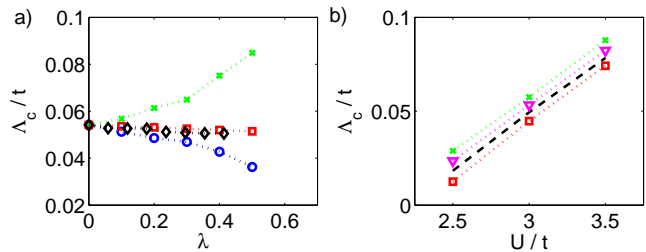}

\caption{(color online). a) Critical scales $\Lambda_c$ for the $d$-wave
  pairing instability vs. electron-phonon coupling $\lambda$, for $T=0.01t$,
  $\mu=-0.95t$, $t'=-t/4$ and $U=3t$. Holstein mode: diamonds), breathing mode:
  squares, $B_{1g}$-mode (Eq. \ref{b1g}): crosses, and apical mode: circles.  
b) Critical scales $\Lambda_c$ for the $d$-wave instability vs. $U$. 
The dashed line is without phonons. Crosses (squares) for only the $B_{1g}$ (apical) mode included, triangles for the breathing,
  $B_{1g}$ and apical mode together.  
All data for $T=0.01t$, $\mu=-0.95t$, $t'=-t/4$.
 }
\label{lambdas}
\end{figure}

Now we include phonon modes. 
The critical scales $\Lambda_c$ for the $d$-wave pairing
instability are shown in Fig. \ref{lambdas} a) as function of the
dimensionless coupling strengths $\lambda$ for the various modes
included separately at $U=3t$.   
The phonon-mediated interaction has two effects which compete. 
First, a momentum-dependent structure can develop which can generate a $d$-wave 
component in the pair scattering, and enhance $d$-wave pairing.  
Second, the attractive part of the retarded interaction reduces the effective onsite repulsion\cite{sangiovanni}, 
which can disfavor spin-fluctuation-induced pairing. The competition between these can be illustrated by the Holstein coupling.

At least for  $U<6t$ it is known that the Holstein coupling $g(\vec{k},\vec{k}')$ is suppressed by the electronic correlations\cite{huang,zeyher}. The suppression is strongest at large $\vec{k}-\vec{k}' \approx (\pi,\pi)$. The fRG for a Holstein mode with frequency 50meV  reproduces this trend\cite{honunpub}. 
In principle, this generates a $d$-wave component in the pair scattering and $\Lambda_c$ for the $d$-wave pairing instability should increase by adding the Holstein phonon. However, the fRG finds a reduction of $\Lambda_c$. The reason is the suppression of the initial Hubbard interaction by the Holstein phonon which outweighs the additional $d$-wave attraction.  

Next we consider the buckling phonon (\ref{buck}) with frequency 50meV. 
Now, already the initial phonon-mediated interaction is attractive for
$d$-wave pairing, as the scattering with  $\vec{k}-\vec{k}' \approx (\pi,\pi)$
is more repulsive than for $\vec{k}-\vec{k}' \approx 0$. But again, the
reduction of the  effective initial repulsion is too strong, and the net
$\Lambda_c$ is lower than for pure electronic interactions. $\Lambda_c$ vs. $\lambda$
behaves similarly to the Holstein case and is not shown in Fig. \ref{lambdas}. We conclude that it is not justified to simply add phonon-mediated pairing interactions on top of unchanged spin-fluctuation-induced interactions in a BCS gap equation. 
The interaction between these two channels needs to be considered. 
Similar trends are found for breathing and the apex oxygen modes. For stronger coupling to the breathing mode, $\lambda_{\mathrm{breathe}} >0.5$, the $d$-wave pairing instability gives way to a $s$-charge-density wave instability with modulation wavevector $(\pi,\pi)$. The only phonon we studied with a positive effect on $\Lambda_c$ for $d$-wave pairing is the $B_{1g}$ buckling mode. 
Since $g_{B1g}(\vec{k},\vec{k}')$ vanishes for $\vec{k}-\vec{k}' \approx (\pi,\pi)$, it supports $d$-wave pairing while it does not suppress the onsite-$U$ due to its $d$-wave form factor. In fact, the fRG energy scale for an AF instability at perfect nesting and half filling is enhanced by the $B_{1g}$-phonon.
In addition, the bare $g_{B1g}$ gets enhanced by the Hubbard interactions\cite{halboth,fu} during the flow. Hence the single $B_{1g}$-mode added to the Hubbard model can increase the energy scale for $d$-wave pairing considerably. For $n=0.83$, $U=2.5t$ and $\lambda_{B_{1g}} = 0.23$\cite{devereaux}, the increase is as high as 55$\%$, for $U=3t$ only 16$\%$. 
With breathing, $B_{1g}$ and apical mode included together (with the $\lambda$s and frequencies cited above, Fig.\ref{lambdas} b)), $\Lambda_c$ is still slightly increased compared to the case without phonons. 
However, such a small increase may be affected by details or uncertainty
  about the $\lambda$s for different phonons.

\begin{figure}

\includegraphics[width=.48\textwidth]{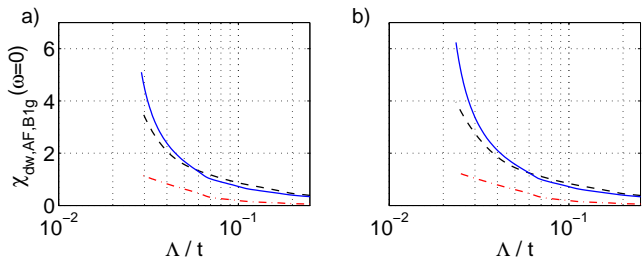}

\caption{(color online). RG flow of $\chi_{dw}$ (solid line), $\chi_{AF}$  (dashed) and
 $d$-wave charge susceptibility at small $q$ (dashed-dotted) for $U=2.5t$, $t'=-0.25t$, $\mu=-0.94t$.  a) with only the $B_{1g}$-mode, $\lambda_{B1g}=0.23$, b) with also apical and breathing modes included, $\lambda_{B1g}=0.23$, $\lambda_{apex}=0.5$, $\lambda_{\mathrm{breathe}}=0.02$.  }
\label{uphon}
\end{figure}

Next we analyze the competition of the $d$-wave superconductor with antiferromagnetism. 
As shown in Fig. \ref{susplot} a), the pairing susceptibility $\chi_{dw}$ overtakes the growth of the AF spin susceptibility $\chi_{AF}$ only very close to the instability. 
When the $B_{1g}$-phonon is included (\ref{uphon} a)), $\Lambda_c$ grows. Also $\chi_{AF}$ is increased at a given scale $\Lambda$, but $\chi_{dw}$ dominates more clearly. This means that the $B_{1g}$ phonon is indeed beneficial for $d$-wave pairing correlations. 
In addition, the coupling to the $B_{1g}$-mode also increases the $d$-wave charge fluctuation tendencies at small wavevectors\cite{fu}. 
This could lead to an additional breaking of the fourfold symmetry of the
FS\cite{halboth,pomeranchuk}. This would not suppress the pairing
instability altogether, but could reduce $\Lambda_c$  by pushing density of states away from the FS.
\\
If we now add the breathing and the apical mode (Fig. \ref{uphon} c)), $\chi_{AF}$ and $\Lambda_c$ get reduced again. However, $\chi_{dw}$ is less affected by the weak breathing mode and the apical mode which barely changes the $d$-wave pairing. Hence for $U=2.5t$, compared to the case without phonons or with the $B_{1g}$ phonon alone, $\chi_{dw}$ dominates even more clearly. Similar trends are seen at $U=3t$. Hence, at least at weak coupling, there is the possibility to enhance $d$-wave pairing in energy scale and with respect to competing instabilities by coupling to the right mix of phonons.   
\\
Finally, commenting on possible caveats for our analysis, we note that selfenergy corrections have been neglected. The reduction of the quasiparticle weight at small energies could potentially reduce the pairing strength even further\cite{dahm}. DCA results\cite{macridin} for $U=8t$ and Holstein phonons are consistent with this. The fRG results give an upper bound for the energy scale for $d$-wave pairing. 
Note that at large $U$ near half filling the coupling to Holstein phonons may actually increase the AF susceptibility\cite{macridin} contrary to our weak coupling results. The reason is the formation of a heavy polaronic quasiparticle which is more effective in an already correlation-narrowed band at large $U$. Nevertheless the decrease of the pairing scale seems a common feature at weak and strong coupling.

In conclusion, we have analyzed the influence of various phonon modes on the $d$-wave pairing instability in the 2D Hubbard model at weak to moderate coupling. Most phonons studied reduce the energy scale for the instability of the Fermi liquid state by reducing the effective onsite repulsion. 
This effect outweighs possible enhancements of the $d$-wave pairing scale due to the wavevector dependence of the electron-phonon coupling. Spin-fluctuation- and phonon-mediated pairing interactions are not additive. 
The only mode studied here which enhances the energy scale for the $d$-wave
instability is the $B_{1g}$ buckling mode.  Due to its
$d$-wave-type wavevector dependence it does not suppress the local onsite
repulsion and therefore does not harm the spin-fluctuation mechanism. For
moderate $U\sim 2.5t$ and average coupling $\lambda_{B1g} = 0.23$, the
increase of the pairing energy scale is more than 50$\%$. 
This increase is reduced when other phonon modes are included. 
Notably, for the parameters used here, where without phonons the $d$-wave pairing and AFM were in very tight competition, the combined effect of three phonons is able to establish the dominance of $d$-wave pairing relative to other instabilities.

CH thanks  D.~J.~Scalapino for stimulating parts of this work. T.~Devereaux, M.~Jarrell, A.~Macridin, W.~Metzner and H.~Yamase are acknowledged for discussions.

\end{document}